\documentstyle[aps,prl,multicol,epsf]{revtex}

\begin{document}
\def\CC{{\rm\kern.24em \vrule width.04em height1.46ex depth-.07ex
\kern-.30em C}}
\def\P{{\rm I\kern-.25em P}}
\def\RR{{\rm
         \vrule width.04em height1.58ex depth-.0ex
         \kern-.04em R}}

\draft
\title{Virtual Quantum Subsystems}
\author{Paolo Zanardi}
\address{
 Institute for Scientific Interchange  (ISI) Foundation,
Viale Settimio Severo 65, I-10133 Torino, Italy\\
Istituto Nazionale per la Fisica della Materia (INFM)
}
\date{\today}
\maketitle
\begin{abstract}
The   physical resources available  to access  and manipulate the degrees of freedom of 
a quantum system define the  set  $\cal A$ of operationally relevant observables.  
The algebraic structure of $\cal A$ selects 
a preferred  tensor product  structure i.e., a partition into subsystems.
The  notion of compoundness for  quantum system  
is accordingly  relativized.
Universal control over virtual subsystems can be achieved by
using quantum noncommutative holonomies
\end{abstract}
\pacs{PACS numbers: 03.67.Lx, 03.65.Fd}

\begin{multicols}{2}

In the last few years  we witnessed a strong
reviving of the interest about the  notion of quantum {\em entanglement} \cite{ent}.
This is mainly due to the essential role that such a concept  is supposed to play
in quantum information processing (QIP) \cite{QC}.
Whenever one has a compounded  (or multi-partite) quantum system,
in the space of admissible states there exist states which
display uniquely quantum  correlations.
These states are referred to as {\em entangled}
and correspond algebraically to the existence, in a
vector space obtained by a tensor product, of vectors $|\psi\rangle$ that are {\em not}
expressible by a simple product e.g., $|\psi_1\rangle\otimes|\psi_2\rangle.$

Given a physical system $S,$
the way to subdivide it in subsystems is in general by no means unique.
On the contrary it is a widespread  praxis in theoretical physics
as well as in everyday life to consider different partitions
into subsystems in dependence of both the physical regime
and the the necessities of the description.
It is is indeed a quite common experience to refer sometimes
to  a system e.g., an atom, as elementary  and sometime as composite
e.g., made out electrons and nucleons. 
The {\em emergence} of a distinguished multi-partite structure
is strongly dependent of the physical regime e.g., the energy-scale,
at which one is working and on the set of observations (experiments) the observer  is interested in.
This is of course a well-known lesson from history of physics e.g., fundamental vs composite particles,
weak-strong coupling dualities, renormalization group etc.

Clearly   even the notion of entanglement is affected by some ambiguity
being {\em relative} to the selected multi-partite structure.
States that are entangled with respect to a given partition in 
subsystems can be separable with respect to another.
Or the other way around: states of a system $S$ that is regarded
as elementary  can be viewed as entangled 
once $S$ is endowed with  a  multi-partite structure. 
In this case  one is in the, somehow paradoxical, situation of having entanglement 
seemingly {\em without} entanglement.

The above ambiguity is removed as soon as, according to some criterion, 
a preferred multi-partite structure is selected among the family
of all possible partition into subsystems.
This selection has in most  cases  a well defined   meaning:
the system $S$ is viewed as composed by $S_1,\, S_2,\ldots$ if  one has
some operational access (is able to "access", "control", "measure") the individual degrees of freedom
of $S_1,\, S_2,\ldots$.
In other terms it is the set of "available" interactions that individuates  the
relevant multi-party decomposition and not an {\em a priori}, god-given partition into elementary subsystems. 
In this letter we shall make an attempt 
towards a formalization of  
the ideas brought about  by these 
 simple remarks.  Our final goal  is to provide a satisfactory algebraic 
definition of what a quantum subsystem is in an operationally motivated framework. 

Let us stress that the notion of {\em virtual} subsystem that we shall introduce 
admits  as a  particular instance the one of quantum code \cite{ERR}, \cite{EAC}
and noiseless subsystems \cite{KLV},\cite{SQI}.
This remark  should make clear  that  virtual subsystems   {\em already} play an important role in QIP.
In particular error avoiding quantum codes  i.e., decoherence-free \cite{EAC}  
 have been   recently also experimentally observed \cite{SCI1}, \cite{SCI2}.

{\em{Compoundness and  tensor products.}}
Let us begin by recalling the basic algebraic structures associated with compoundness.
Let $S_1$ and $S_2$  be two {\em classical} systems  with configuration manifolds
${ M}_i\,(i=1,2).$ Roughly speaking the associated quantum systems
have state-spaces given by ${\cal H}_i= {\cal F}(M_i)$
where ${\cal F}$ denotes some suitable (complex-valued) function space over the $M_i$'s 
e.g., $L^2$-summable functions.
Notice that  these spaces (actually {\em abelian} $C^*$-algebras \cite{ALG}) are
the classical ``observable'' spaces; the quantum ones are given by the operator
(non-abelian) algebras End$({\cal H}_i).$
In the classical realm the manifold associated with  the joint systems $S_1\vee S_2$
is given by the {\em cartesian} product $M_1\times M_2.$
It follows that at the {\em quantum} level one has ${\cal H}_{S_1\vee S_2}={\cal H}_1\otimes {\cal H}_2$
indeed 
 ${\cal F}(M_1\times M_2)$ is given by a suitable closure of ${{\cal F}(M_1)\otimes {\cal F}(M_2)}.$
This  basic  functorial identity is the {\em algebraic} ground for the quantum theory axiom
associating to a bi-partite systems a state-space given by the tensor products
of the state-spaces describing the {\em subsystems}.
The extension to $N$-partite systems is obvious.
One has another elementary, yet  remarkable, functorial relation
given by the  {\em canonical } isomorphism
$\mbox{End}({\cal H}_1\otimes {\cal H}_2)\cong \mbox{End}({\cal H}_1)\otimes\mbox{End}({\cal H}_2).
$
Even in  the quantum realm the observable algebra associated with a joint systems
is given by the tensor product of the subsystem subalgebras.

Our  key observation is that different kinds of compoundness can emerge in the same system
when one considers different sets of observables as the physical ones.
Indeed quite often it  makes    sense  to refer to a subalgebra $\cal A$
(rather than the full operator algebra)
as to the {\em physical} observable algebra. Limitations of physical resources
may   lead  to select a specific class of operators to be  considered as realizable.
For instance energy supply limitations  lead naturally to restrict to
 operators $X$  which have  vanishing matrix elements between energy eigenstates
whose energy difference exceeds some bound $E.$
At the dynamical-algebraic level the selection of a particular multi-party decomposition  means that the
algebra of (operationally relevant) observables ${\cal A}$
has a tensor product structure (TPS) i.e,  ${\cal A}\cong\otimes {\cal A}_i$
such that all the observables belonging to the individual  ${\cal A}_i'$s
can be effectively implemented.

Before passing  to general constructions 
it is useful to consider a very simple example
in which one has a set  of subsystems (degrees of freedom) associated with 
a rapidly growing sequence of energy scales. Starting from the ground state 
and increasing the energy available one is able to excite more and more subsystems 
that at lower energy were frozen.
This situation is realized, for instance, in systems in which one has confined directions
or in the cases in which an adiabatic decoupling between fast and slow degrees of freedom has been performed:
the effective dimensionality
of the system is a function of the energy scale.

{\em  {The TPS manifold}}.
Let us consider an Hilbert space ${\cal H}\cong \CC^n$ with a priori {\em no} tensor product structure.
A first very natural question is
 {\em how many inequivalent TPS's can be assigned over $\cal H$?}
Or more physically: in how many different ways ${\cal H}$ can be viewed as the state-space of a multipartite
quantum system?
If $n$ is a prime number there are no possibilities: the system is {\em elementary}.
If $n$ is {\em not} prime it has a non-trivial prime factorization:
$n=\prod_{i=1}^r p_i^{n_i}\,(p_i< p_{i+1}).$
If the exponent $n_i$ of the $i$-th prime factor of $n$
is not one then several regroupings are possible e.g., $r=1,\,p_1=2,\, n_1=3\Rightarrow
3=1+1+1,\,3=1+2$ corresponding to the state-space factorizations
$\CC^8\cong\CC^2\otimes\CC^2\otimes\CC^2$ and $ \CC^8\cong \CC^2\otimes \CC^4.$
When more than one $p_i$ appear in the decomposition of $n$
we see that many other possibilities of writing $n$ as a product of integers
arise. In general, given $n,$ we introduce the set of factorizations
${\cal P}_n=\{P\subset {\bf{N}}\,/\, \prod_{m\in P} m=n \}
$
where $ {\bf{N}}$ denotes the set of natural numbers.

Given  a factorization  $P=\{ n_1\le n_2\le\ldots\le n_{|P|}\}\in{\cal P}_n$ of $n$ is assigned
one has the (non-canonical) isomorphisms 
$\varphi\colon {\cal H} \mapsto  \otimes_{j=1}^{|P|} \CC^{n_j}.
$
In the following such isomorphisms will be referred  to as {\em tensor product structures} (TPS)
 over $\cal H,$ and subsystems of the associated
multi-party decomposition  as {virtual}.

Given a distinguished  TPS , say $\varphi_0,$
one can identify the group of unitaries
${\cal U}({\cal H})$ and ${\cal U}(\otimes_{j}\CC^{n_{j}})$
via the algebra isomorphism $U\mapsto \varphi^{-1}_0\circ U\circ \varphi_0$.
A suitable quotient of this latter unitary group parametrizes the space of inequivalent TPS's.
Indeed two elements $U$ and $W$ of ${\cal U}(\otimes_{j}\CC^{n_{j}})$
define equivalent TPS's if either    $U=U_1\,W\,U_2$ where the $U_i$'s are {\em multi-local}
transformation i.e., $U_i\in\prod_{k=1}^{|P|} U(n_k),\,(i=1,2);$  or
the $U_i$s are {\em permutations} of  factors with equal dimension.
In the first case the TPS's  differ just by  a change of the basis in each factor, in the second
by the order of the factors that in turn amounts simply
to a  relabelling of the subsystems. The space of {\em inequivalent}  TPS's over $\CC^n$
will be denoted by ${\cal T}_n.$

Once a given multiplicative partition $(n_i)$ of $n$ is chosen  along with a particular $\varphi$ one has
${\cal H}=\otimes_{i=1}^N {\cal H}_i ,\,({\cal H}_i:=\varphi( \CC^{n_i}))$
then End$({\cal H})\cong\otimes_{i=1}^N {\cal A}_i$ where
${\cal A}_i:=\mbox{End}({\cal H}_i).$
For any  set of unitaries in $\cal H$ labelled by the elements $\lambda$
of some manifold $\cal M,$ e.g., external fields,
one can define  ${\cal A}_i(\lambda):=U_\lambda\,{\cal A}_i\,U^\dagger_\lambda, (i=1,\ldots,N)$
that describes  a family of multipartite structures over ${\cal H}$ parametrized
by points of  $\cal M$. As noticed above not all the points of $\cal M$
necessarily correspond to different TPS's. Indeed
it can happen  that different $\lambda$'s can result in the same structure
e.g., $U_\lambda\,{\cal A}_i\,U^\dagger_\lambda={\cal A}_i.$
If a state is entangled (product) with respect to a TPS labelled by $\lambda\in{\cal T}_n$
it will be referred to as $\lambda$-entangled ($\lambda$-product).

If $E\colon {\cal H}\mapsto \RR^+_0$ denotes an entanglement measure over $\cal H$
with respect to  a given TPS, say $\lambda=0,$  one has that $E_\lambda:= E\circ U_\lambda$
is a  $\lambda$-entanglement measure. 
In turn the latter provides a natural measure of the "distance" between the TPS at  $\lambda\neq 0$
and that  at $\lambda=0.$
Indeed it appears quite natural to say that the more the $\lambda=0$  product states 
are $\lambda$-entangled  the more the TPS at $\lambda$ differs from the one at the origin.
To make this idea quantitative one has to make it independent on the particular 
state; this can be done either by maximizing or by taking 
the  average over all the $0$-product states.
In this latter case one finds that the distance  one is  looking for
is nothing but the (square root of)  {\em entangling power} of $U_\lambda$ \cite{ep}:
$e(U)=\int d\psi_1 d\psi_2\,{E(U\,|\psi_1\rangle\otimes |\psi\rangle_2)}.$
Here the integral is done with respect to the uniform e.g., Haar, measure
over the pure product state manifold.

In order to exemplify 
the notion of TPS manifold of  we now introduce  a  family of TPS's over an infinite dimensional state-space
parametrized by a group of $N\times N$ matrices. 
Let us consider $N$ harmonic oscillators. The global state-space
is given by ${\cal H}_N:= \otimes_{i=1}^N {\cal H}_i$ where each of the factors
is the single boson Fock space i.e., ${\cal H}_i= \mbox{span}\,\{|n\rangle\}_{n\in{\bf{N}}}$
associated with the annihilation and creation operators $a_i$ and $a_i^\dagger,\,
(a_i^\dagger\,a_i|n\rangle=n\,|n\rangle).$
Let $U\in{ U}(N)$ be a  complex $N\times N$ unitary matrix.
The operators 
$
a^U_i:= \sum_{j=1}^N U_{ji} \,a_j\quad (i=1,\ldots,N)
$
represents new bosonic {\em modes} 
i.e., $[a_i^U,\,a_j^{U\dagger}]=\delta_{ij},\,[a_i^U,\,a_j^U]=0,$
moreover one has ${\cal H} =  \otimes_{j=1}^N  {\cal H}_i^U$
where the ${\cal H}_j^U$'s are the Fock spaces associated with   the ${a}^U_j$'s.
Notice that the Fock vacuum $|0\rangle :=\otimes_i |0\rangle_i$ is $U$ independent
i.e., $a_J^U\,|0\rangle=0 (\forall U,\,j)$.
One has ${\cal H}_j^U\cong {\cal A}_j^U\,|0\rangle$ where  ${\cal A}_j^U$
is the algebra generated by $a_j^U$ and $a_j^{U\dagger}.$  
States like $a^{U\dagger}_j\,|0\rangle$ are unentangled with respect to the TPS 
defined by the given $U$ but entangled with respect to the one associated with e.g., $U=\openone.$

{\em Virtual bi-partitions.}
Now we address the following issue:
when is it legitimate to consider a pair of observable algebras
as describing a bi-partite quantum system?
Suppose that ${\cal A}_1$ and ${\cal A}_2 $
are two {\em commuting} $*$-subalgebras of ${\cal A}:=\mbox{End}({\cal H}) $
such that the subalgebra  ${\cal A}_1\vee{\cal A}_2 $
 they generate 
i.e., the minimal  $*$-subalgebra containing both
${\cal A}_1$ and  ${\cal A}_2,$
amounts to the whole ${\cal A}$
and moreover one has the (non-canonical) algebra isomorphism
\begin{equation}
{\cal A}_1\vee{\cal A}_2\cong {\cal A}_1\otimes {\cal A}_2.
\label{ss-alg}
\end{equation}
The standard, {\em genuinely } bi-partite, situation is of course ${\cal H}={\cal H}_1\otimes {\cal H}_2,\,
{\cal A}_1=\mbox{End}({\cal H}_1)\otimes\openone,\,{\cal A}_2=\openone\otimes\mbox{End}({\cal H}_2).$
If ${\cal A}_1^\prime:=\{ X\,/\, [X,\,{\cal A}_1]=0 \}$ 
denotes the {\em commutant} of ${\cal A}_1,$ in this case one  has 
that ${\cal A}_1^\prime={\cal A}_2.$

It is important to  mention that a prototypical and ubiquitous situation described by Eq. (\ref{ss-alg})
is when ${\cal A}_1$ and ${\cal A}_2$ are {\em local} observable algebras
associated to { disjoint} regions of space at equal time.
More generally such an independence of local degrees of freedom
e.g., quantum fields, is encoded in terms of  commutativity 
between observables supported on causally disconnected domains \cite{haag}.  
Notice also the spatial separation between parties 
e.g., Alice and Bob, is a common assumption in  
protocols for quantum communication e.g, teleportation \cite{QC}.

The point of view 
advocated in this letter is  to consider condition (\ref{ss-alg}) as the {\em definition} of bi-partite system.
regardless the "real" compoundness or not of the  underlying state-space.
Accordingly we shall  consider as "real" entanglement the one  occurring in that case.
The (nearly obvious) point is that:
in order to take computational advantage from this virtual
entanglement one
must have {\em access} to i.e., to be able to control  the subalgebras ${\cal A}_{1, 2}.$
As far as the  operations in ${\cal A}_1$ and ${\cal A}_2$
are easily realizable (accessible) in the lab we shall consider them as primitive and local,
regardless how they look at the original level.

The theory of {\em noiseless subsystems} \cite{KLV}, \cite{SQI}
provides an important exemplification
as well as
source of inspiration for  the  approach to compoundness
advocated here.
Let us consider a system made of $N$ "real" subsystems e.g., qubits.
Suppose that the algebra  of relevant  interactions
is given by ${\cal A}_1\cup{\cal A}_1^\prime$  where ${\cal A}_1$  \cite{ALG}
\begin{equation}
{\cal A}_1\cong\oplus _J \openone_{n_J}\otimes M_{d_J}( \CC).
\label{central}
\end{equation}
This decomposition reads at the state-space level as ${\cal H}\cong \oplus_J \CC^{n_J}\otimes \CC^{d_J}.$
For a fixed  label  $J$  one has that the elements
of ${\cal A}_1$ (${\cal A}_1^\prime$) act as the identity on the $\CC^{n_J}$  ($\CC^{d_J}$) factor.
This means that the system is viewed, for all practical purposes, as a bipartite
one, in which the observables of the first (second) subsystems are given by ${\cal A}_1$ (${\cal A}_1^\prime$).
For collective decoherence, ${\cal A}_1$ is the interaction
algebra generated by couplings with the environment
invariant under qubit permutations. While ${\cal A}_1^\prime$
is given by any linear combinations of permutation operators \cite{SQI}.
In particular the  latter algebra is generated by
exchange i.e., Heisenberg like operators between the different pairs of qubits \cite{divi-nature}.

Generally speaking  Eq. (\ref{central}) shows in which sense an observable  algebra ${\cal A}_1$
(${\cal A}_1^\prime$)
is associated  with a collection  of virtual subsystems, i.e., the $\CC^{d_J}$ ($\CC^{n_J}$) factors,
 labelled by  its spectrum. 
It is worth  to observe that when ${\cal A}_1$ is {\em abelian} all the $d_J$'s are equal to one.
In this case, if $n_J>1,$ the $J$-th factor of the state-space decomposition describes a sort of hybrid
bi-partite system  in which one of the factors  is quantum whereas  the other 
represents  a classical system with a one-point configuration space.
This is exactly the situation one meets in the case of quantum codes, both
error correcting   \cite{ERR}   and error avoiding \cite{EAC}.
In this latter case the algebra ${\cal A}_1$ is generated by the operators
coupling the computing system with its environment
and  ${\cal A}_1^\prime$ is the set of interactions necessary to perform computations
entirely within the decoherence-free sector \cite{SQI}.

To make clear the connection with quantum error correction
let us consider  a set $\{ X_i\}_{i=1}^k$  of $k\le n$ linear independent traceless "parity" operators over
${\cal H}\cong (\CC^2)^{\otimes\,n},$ such that 
$X_i=X_i^\dagger,\, X_i^2=\openone,\, [X_i,\,X_j]=0,
(i,j=1,\ldots,k).$
Following  standard arguments of quantum error correction \cite{ERR}
one can show that the $X_i$'s   generate an abelian algebra ${\cal A} \cong \CC {\bf{Z}}_2^k.$
The   associated state-space decomposition is given by
\begin{equation}
{\cal H}\cong \oplus_{J\in{\bf{Z}}_2^k} \CC^{2^{n-k}}\otimes\CC\cong  \CC^{2^{n-k}}\otimes\CC^{2^k}.
\end{equation}
It easy to see that the commutant of $\cal A$ contains the algebra
of operators over  the first  factor in the decomposition above
This means that the  set of operators with well-defined parities
defines  and controls a virtual subsystem of $n-k$ bits.
Analogously the set of "odd" operators   ($\{ O\,/\,\exists i\,
\{X_i,\,O\}=0\}$)
defines and controls the second  $k$-qubit subsystem.
For instance   the parity $X_1:=\sigma_x\otimes\openone$
defines the natural  bi-partite  structure over $(\CC^2)^{\otimes\,2}$
whereas $X_1^\prime=\sigma_x^{\otimes\,2}$ defines 
TPS such that states like  $2^{-1/2} (|00\rangle\pm |11\rangle)$
are  unentangled.
Notice that in errror correction theory the first (second) subsystem is related to the code (syndrome).
For any unitary $U,$ the operators $X_i(U):= U\,X_i\,U^\dagger$
span an algebra isomorphic to  $\cal A$ above.
Again one has a continuous set of TPS's parametrized by points  of a unitary group \cite{illustra}.

Turning back to the  characterization of pairs of (finite-dimensional) subalgebras satisfying 
Eq. (\ref{ss-alg}) by using Eq. (\ref{central}) it is easy to prove the following \cite{proof}

{\em {Proposition}}
 Let ${\cal A}_1$ and ${\cal A}_2$ be two commuting  $*$-subalgebras of a  finite dimensional
$*$-algebra ${\cal A}.$
Necessary and sufficient condition for the validity of (\ref{ss-alg}) is that  
  ${\cal A}_1\cap {\cal A}_1^\prime=\CC\,\openone$ i.e., ${\cal A}_1$ is a factor.

{\em Holonomic control on subsystems.}
In this paragraph we  show that the Holonomic approach to  QC
\cite{HQC} provides a natural setting for the issue
of information processing within a (virtual) subsystem.

Let $X\in \mbox{End}({\cal H})\cong M_{n d}(\CC)$ be 
an hermitean operator  with  a spectrum of  $d$ iso-degenerate eigenvalues 
i.e., $X=\sum_{i=1}^d x_i \sum_{k=1}^n |k i\rangle\langle k i|,$
and $\{U_\lambda\}_{\lambda\in{\cal M}}\subset U({\cal H})$ a set of unitaries 
parametrized by the point of some (control) manifold ${\cal M}.$
Then the set of $X(\lambda) := U_\lambda\, X\,U_\lambda^\dagger,
$ is a family that in the generic case, for sufficiently large $D=$dim ${\cal M}$
satisfies  the conditions for (universal) holonomic quantum computation \cite{HQC}
on  the $n$-dimensional degenerate eigenspace ${\cal C}_i= \mbox{span}\,
\{|k i\rangle\}_{k=1}^n \cong \CC^n\otimes |i\rangle,\,(i=1,\ldots,d)$
of $\openone_n\otimes X.$   
This implies  that the {\em holonomy group} Hol$(A_i)$ associated with the connection $u(n)$-valued 1-forms
$A_i^{ab}= \langle a| \otimes\langle i| \,U_\lambda^\dagger\,d\, U_\lambda\, |b\rangle\otimes|i\rangle,\,
d:=\sum_{\mu=1}^D d\lambda_\mu\, \partial_\mu,\, (a,b=1,\ldots,n)$
is the whole $U({\cal C}_i)\cong U(n)\otimes |i\rangle\langle i|$ \cite{HQC}.
By denoting collectively with $A$ the set of the $A_i$'s one can therefore
write that
\begin{equation} 
\mbox{Hol}(A)\cong 
 \oplus_{i=1}^d U(n)\otimes|i\rangle\langle i|
\supset U(n)\otimes \openone_d.
\nonumber
\end{equation}
The last inclusion tells us in that in the generic case the holonomy group of $A$
will contain  the whole  unitary group of the 
$\CC^n$ subsystem. 
Once the holonomic family $\{X(\lambda)\}_\lambda$ is given, 
any transformation i.e., computations in the first subsystem
can be generated holonomies. 
Notice that, since for real quantum case one must have $n\ge 2,$
the holonomy group is necessarily nonabelian.

{\em Conclusions.}
We analyzed some the consequences of 
the non uniqueness of the decomposition  of a given system $S$
into subsystems.
Such non-uniqueness  implies,  at the quantum level, a  fundamental ambiguity about the 
very notion of entanglement that accordingly becomes a {\em relative} one.
One can parametrize the space of all possible partitions i.e, tensor product structures,
of a $n$-dimensional quantum state-space by  the points of a set  ${\cal T}_n.$
The fact of considering  all the points in ${\cal T}_n$ on the same footing
(that amounts to establishing a   {\em democracy} between different TPS's)
 provide a  relativization of the  notion of entanglement.
Without further physical assumption,
no partition has an ontologically superior status with respect to any other.
The subsystems associated with all these possible i.e, {\em potential}
multi-party decomposition were referred to as  virtual. 
A distinguished point of ${\cal T}_n$ is selected i.e., made {\em actual} only once the relevant algebra 
$\cal A$ of "physical" observables
is given. Indeed considering a given partition as the privileged 
 has a strong  { operational} meaning, in that it  depends on
the set of  resources  effectively available to access and to control the degrees
of freedom of $S.$
Different sets of resources give rise to  different partitions physically relevant.
We provided several examples of natural, though hidden, multipartite structures 
arising from the given   algebraic structure of $\cal A.$
We briefly showed that the holonomic approach to 
quantum computation provides one  natural way to address
the issue of controllability  within virtual subsystems.
We believe   that this democratic approach to quantum compoundness  is, on the one hand, sound
from the conceptual point of view, and on the other  hand possibly relevant to QIP.

The author acknowledges  R. R. Zapatrin and M. Rasetti for stimulating discussions
and encouragement.

\end{multicols}

\begin{references}
\bibitem{illustra} Let us consider  two qubits
and the  family  by $U_\lambda:=\exp(i\,\lambda\, S)= \cos \lambda \,\openone +
i\,\sin \lambda\,S$
where  $S\,|\psi\rangle\otimes|\phi\rangle= |\phi\rangle\otimes|\psi$).
If $X_1:=\sigma_x\otimes\openone$  one has
$X_1(U_\lambda) =\cos^2\lambda\, X_1
+ \sin^2\lambda\,\openone\otimes\sigma_x +
{i}/{2}\,\sin 2\lambda\,[S,\,X_1].$
Clearly  $X_1(\pi/2)= \openone\otimes\sigma_x:$
the TPS associated with $\lambda=\pi/2$ amounts just to the exchange
between the two subsystems.
 All the points of $[0,\,\pi/2)$ correspond to inequivalent TPSs.
\bibitem{proof}
We can assume ${\cal A}_2={\cal A}_1^\prime.$
If ${\cal A}_1$ is a factor  one has ${\cal A}_1\cong \openone_{n}\otimes M_d( \CC)$ and then
 ${\cal A}_2= M_n( \CC)\otimes  \openone_n.$
There  follows that ${\cal A}_1\vee{\cal A}_2\cong  M_n( \CC)\otimes M_d( \CC)={\cal A}_1\otimes {\cal A}_2.$
When ${\cal A}_1$ is {\em not} Eq. (\ref{central})
 implies ${\cal A}_2\subset \oplus _J  M_{n_J}( \CC)\otimes\openone_{d_J}$
then ${\cal A}_1\vee  {\cal A}_2=  \oplus_J  M_{n_J}( \CC)\otimes M_{d_J}(\CC).$
A comparison  of the dimension of this latter algebra 
with that of  ${\cal A}_1\otimes  {\cal A}_2$ completes the proof.
\end{references}
\end{document}